\documentclass[twocolumn]{aastex631}

\usepackage{natbib}
\usepackage{amsmath}
\usepackage{braket}
\usepackage{color}
\hypersetup{
 colorlinks=true,%
 linkcolor=black,
 citecolor=blue,
}

\begin{document}
\def\vec#1{\mbox{\boldmath $#1$}}
\newcommand{\average}[1]{\ensuremath{\langle#1\rangle} }

\title{\large{\bf Testing the Alfv\'en-wave model of the solar wind with interplanetary scintillation
}}

	\author{Munehito Shoda}
	\affiliation{National Astronomical Observatory of Japan, National Institutes of Natural Sciences, 2-21-1 Osawa, Mitaka, Tokyo, 181-8588, Japan}
	\author{Kazumasa Iwai}
	\affiliation{Institute for Space-Earth Environmental Research, Nagoya University, Furo-cho, Chikusa-ku, Nagoya, 464-8601, Japan}
	\author{Daikou Shiota}
	\affiliation{Institute for Space-Earth Environmental Research, Nagoya University, Furo-cho, Chikusa-ku, Nagoya, 464-8601, Japan}
	\affiliation{National Institute of Information and Communications Technology (NICT), Nukui-Kita-machi, Koganei,Tokyo, 184-8795 Japan}

	\correspondingauthor{Munehito Shoda}
	\email{munehito.shoda@nao.ac.jp}

\begin{abstract}

  Understanding the mechanism(s) of the solar wind acceleration is important in astrophysics and geophysics.
  A promising model of the solar wind acceleration is known as the wave/turbulence-driven (WTD) model, 
  in which Alfv\'en waves feed energy to the solar wind.
  In this study, we tested the WTD model with global measurement of wind speed from interplanetary scintillation (IPS) observations.
  For Carrington rotations in minimal and maximal activity phases,
  we selected field lines calculated by the potential-field source-surface method in high- and mid-latitudes and compared the simulated and observed wind velocities.
  The simulation was performed in a self-consistent manner by solving the magnetohydrodynamic equations from the photosphere to the solar wind.
  In high-latitude regions, the simulated solar wind velocity agrees better with the IPS observation than with the classical Wang--Sheeley empirical estimation, both in maximal and minimal activity phases.
  In mid-latitude regions, the agreement worsens, possibly because of the inaccuracy of the WTD model and/or the magnetic-field extrapolation.
  Our results indicate that the high-latitude solar wind is likely to be driven by waves and turbulence,
  and that the physics-based prediction of the solar wind velocity is highly feasible with an improved magnetic-field extrapolation.

\end{abstract}


\section{Introduction} \label{sec:introduction}

The solar wind \citep{Parker_1958_ApJ,Velli_1994_ApJ} plays several important roles in astrophysics and heliophysics. 
It is the principal source of solar angular-momentum loss \citep{Weber_1967_ApJ,Sakurai_1985_AA,Kawaler_1988_ApJ,Reville_2015_ApJ,Finley_2019_ApJ,Finley_2020_ApJ}, which results in solar spin-down \citep{Irwin_2009_proceedings,Gallet_2013_AA,Matt_2015_ApJ} and weakening of solar magnetic activity over time \citep{Gudel_2007_LRSP,Brun_2017_LRSP,Vidotto_2021_LRSP}.
In the space weather forecasting, 
it is necessary to predict the velocity and spatial structure of the solar wind (including impulsive eruptions) that affect the formation of the co-rotating interaction region \citep[CIR,][]{Smith_1976_GRL,Tsurutani_2006_JGRA} and the arrival time of the coronal mass ejection \citep[CME,][]{Shiota_2016_SpaceWeather,Wold_2018_JSWSC,Riley_2021_SpaceWeather}.
Fluctuations in the solar wind are also important for studying plasma turbulence that extends from the fluid to electron scales \citep{Bruno_2013_LRSP}.

One fundamental open question of the solar wind is its heating and acceleration mechanism(s).
The classical thermally driven wind model of the solar wind \citep{Parker_1958_ApJ,Withbroe_1988_ApJ,Hansteen_1995_JGR} does not account for the observed anti-correlation between the coronal temperature and wind velocity \citep{Geiss_1995_SSRev,von_Steiger_2000_JGR}, 
and thus an additional non-thermal acceleration of the solar wind is necessary \citep{Alazraki_1971_AA,Belcher_1971_ApJ,Jacques_1977_ApJ}.
In terms of feeding mechanism of the magnetic energy, the solar wind models are classified into two types \citep{Cranmer_2012_SSRev}: wave/turbulence-driven (WTD) models \citep{Suzuki_2005_ApJ,Cranmer_2007_ApJ,Verdini_2010_ApJ,van_der_Holst_2014_ApJ,Reville_2020_ApJ} and reconnection/loop-opening (RLO) models \citep{Fisk_2003_JGR,Antiochos_2011_ApJ,Higginson_2017_ApJ}.

\begin{figure*}[t!]
\centering
\hspace{1em} \includegraphics[width=160mm]{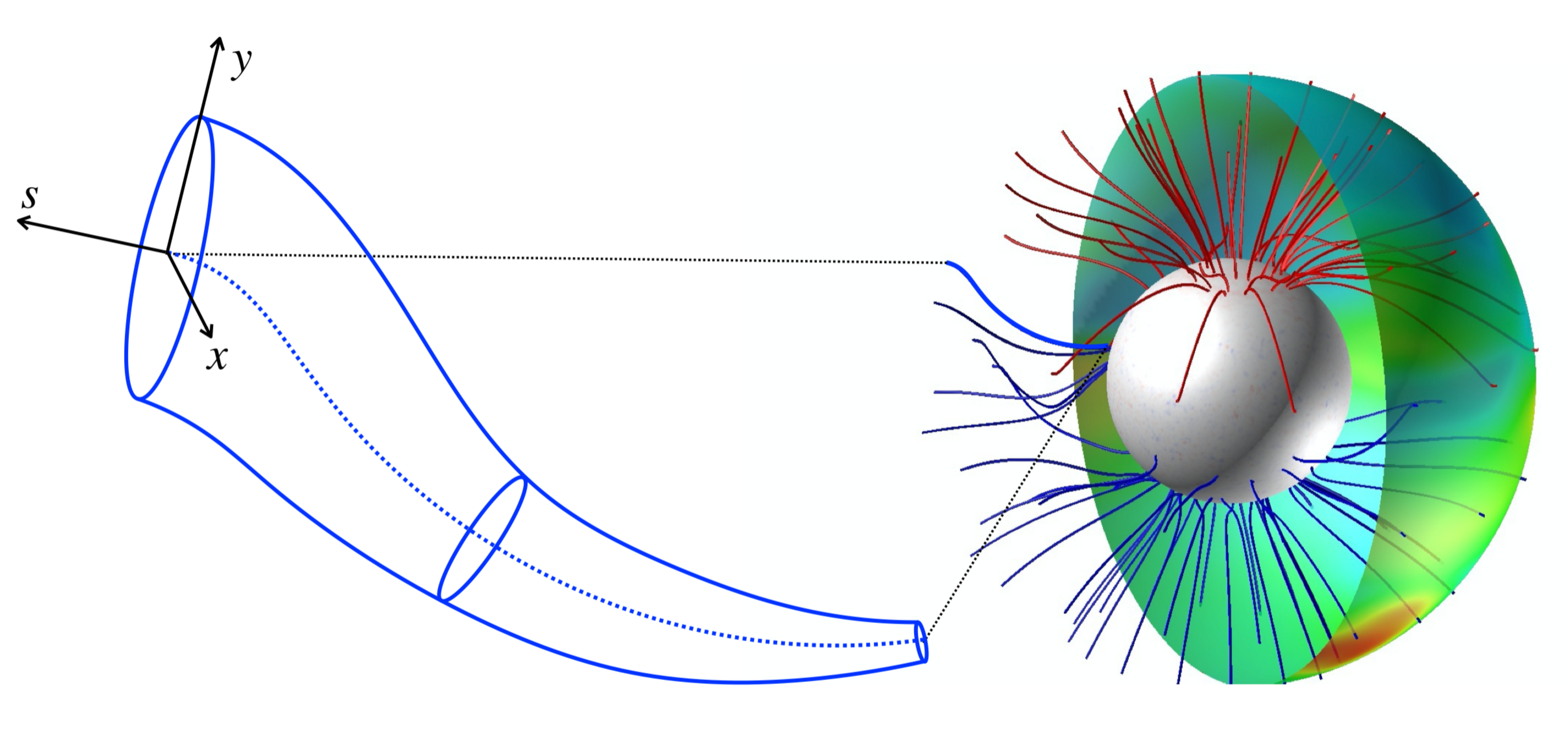}
\vspace{1em}
\caption{
Model overview. 
Right: 3D view of the photospheric magnetic field measured by GONG and extrapolated field lines with color indicating the polarity. 
Also shown by the transparent hemisphere is the corresponding IPS measurements of the solar wind velocity.
Left: a schematic of the simulation domain.
For a given magnetic field line (flux tube), we solve the magnetohydrodynamic equations along the field-aligned coordinate $s$.
}
\label{fig:overview}
\vspace{1em}
\end{figure*}

Several theoretical studies have indicated that the reconnection/loop-opening processes have minor roles in coronal holes \citep{Cranmer_2010_ApJ,Lionello_2016_ApJ}.
Meanwhile, a recent observational study indicated a non-negligible role of emerging flux (that induces reconnection and loop opening) in the coronal-hole wind \citep{Wang_2020_ApJ}.
Models of magnetic switchbacks \citep{Bale_2019_Nature,Kasper_2019_Nature,Dudok_de_Wit_2020_ApJ,Horbury_2020_ApJ} observed by Parker Solar Probe \citep{Fox_2016_SSRev} also suggest that interchange reconnection frequently occurs at the base of the coronal hole \citep{Sterling_2020_ApJ,Zank_2020_ApJ,Drake_2021_AA,Magyar_2021_ApJ}, although it should be noted that switchbacks can emerge without interchange reconnection \citep{Squire_2020_ApJ,Schwadron_2021_ApJ,Shoda_2021_ApJ}.
The acceleration mechanism(s) of the solar wind, even those from coronal holes, is still (or getting more) controversial.

In this study, we test the wave/turbulence model of the solar wind using interplanetary scintillation (IPS) observations \citep{Hewish_1964_Nature,Salpeter_1967_ApJ},
which yields the global structure of the solar wind velocity \citep{Kojima_1998_JGR,Tokumaru_2009_GRL,Porowski_2021_arXiv}.
For this purpose, we perform one-dimensional solar wind simulations for multiple open flux tubes and compare the simulated solar wind velocity with the IPS measurements.
An advantage of our model over the previous multiple flux-tube model of the solar wind \citep[MULTI-VP,][]{Pinto_2017_ApJ} is that we model the solar wind with Alfv\'en-wave heating and acceleration mechanisms,
while MULTI-VP is more practical in the space weather forecasting because of its small computational cost \citep{Samara_2021_AA}.
The radial structure of the open magnetic field is required as an input for the simulation and is inferred from the potential-field-source-surface (PFSS) extrapolation of the photospheric magnetic field.
In particular, we investigate how accurately the wave/turbulence model predicts the solar wind velocity for a given magnetic field configuration.

The reminder of this paper is organized as follows.
The extrapolation of the magnetic field and simulation setting are detailed in Section 2.
In Section 3, we first provide an overview of the simulation results and thereafter compare them with the IPS measurements for the activity-minimum and -maximum phases.
Finally, the discovery and possible future improvements are discussed in Section 4.

\section{Model} \label{sec:model}

\subsection{Model overview}

The overview of the model is shown in Figure \ref{fig:overview}.
The large-scale magnetic field was extrapolated using the PFSS method \citep{Altschuler_1969_SolPhys,Schatten_1969_SolPhys,Shiota_2012_ASPC}.
Given the shape (cross-section) and inclination (angle with respect to the vertical direction) of the flux tube,
one-dimensional magnetohydrodynamic equations were numerically solved along the flux tube from the photosphere to the solar wind.
The simulated velocity was thereafter compared to the IPS observations to test the model accuracy.
We describe the details of the magnetic-field extrapolation and basic equations in Sections \ref{sec:background_magnetic_field} and \ref{sec:basic_equations}, respectively.

\subsection{Background magnetic field \label{sec:background_magnetic_field}}

To perform a one-dimensional solar wind simulation, 
we need to prescribe the field strength $B_s$ and the inclination with respect to the vertical (radial) direction $\theta_{BR}$ of the background flux tube.
The detailed procedure is described as follows.

First, we extrapolate the magnetic field using the PFSS method.
The source surface radius $r_{\rm SS}$ is set to the conventional value of $r_{\rm SS}/R_\odot = 2.5$ \citep{Hoeksema_1983_JGR,Wang_1988_JGR}.
The GONG synoptic map \citep{Harvey_1996_Science} is used as the photospheric boundary condition.
For simplicity, we ignore the effect of solar rotation,
which is unlikely to affect the solar wind velocity \citep{Weber_1967_ApJ}.
For a given set of latitude and longitude at the source surface, we obtained the corresponding field-aligned profiles of the radial distance $r^{\rm PFSS}$ and field strength $B^{\rm PFSS}$.
It is, however, risky to directly adopt the magnetic field extrapolated in this manner.
The open magnetic flux obtained from the PFSS method ($r_{\rm SS}/R_\odot = 2.5$) with the GONG magnetogram was found to be three time smaller than the observed value \citep[open-flux problem, see][]{Linker_2017_ApJ,Linker_2021_arXiv}.
This gap is attributed to the missing magnetic flux in the GONG magnetogram and/or the inaccuracy in the PFSS extrapolation.
To make the open flux consistent with the in-situ observations, 
the background magnetic field is defined by the PFSS field multiplied by three:
\begin{align}
    B_s = 3 B^{\rm PFSS} \left( s \right), \ \ \ \ r = r^{\rm PFSS} \left( s \right),
    \label{eq:magetic_field_upper}
\end{align}
where $B_s$ denotes the field strength of the background flux tube, $r$ is the radial distance and $s$ is the field-aligned coordinate.
Because the PFSS extrapolation is incapable of describing the magnetic field in the lower atmosphere,
Eq. (\ref{eq:magetic_field_upper}) is adopted for $r/R_\odot>1.02$.

In the lower atmosphere ($1 \le r/R_\odot \le 1.02$), although direct observation of the magnetic field is still difficult, 
flux tubes appear to expand with height in response to an exponential decrease in the ambient gas pressure \citep{Ishikawa_2021_ScienceAdvances}.
Here, we simply assume that the magnetic field expands to keep the plasma beta nearly constant until the field strength reaches the coronal-base value.
\begin{align}
    B_s \left( s \right) = \max \left[ B_{s,{\rm cb}}, B_{s,\odot} \exp 
    \left( - \frac{r -R_\odot}{2 H_{p,\odot}}  \right) \right], \label{eq:magetic_field_lower}
\end{align}
where $B_{s,{\rm cb}}$ denotes the field strength at the coronal base,
$B_{s,\odot}$ is the photospheric field strength, 
and $H_{p,\odot} = 150 {\rm \ km}$ is the pressure scale height in the photosphere.
Assuming that the pressure scale height is constant in the photosphere and chromosphere, the height distribution of the pressure is given as follows:
\begin{align}
    p(s) \approx p_\odot \exp \left( - \frac{r -R_\odot}{H_{p,\odot}} \right),
\end{align}
yielding the plasma beta constant in $s$:
\begin{align}
    \beta(s) = \frac{8 \pi p(s)}{B_s(s)^2} \approx \frac{8 \pi p_\odot}{B_{s,\odot}^2} \ \ \ ({\rm if} \ B_s > B_{s,{\rm cb}}).
\end{align}
Although the actual chromospheric magnetic field could deviate from Eq.~\eqref{eq:magetic_field_lower}, we have confirmed that the solar wind velocity is only slightly affected by the formulation of $B_s$ in the lower atmosphere (see Appendix for detail).
Therefore, the arbitrary choice of $B_s$ does not significantly affect the conclusion of this work.

We require that the background field strength is continuous.
From Eq. (\ref{eq:magetic_field_upper}) adopted in $r/R_\odot>1.02$, and Eq. (\ref{eq:magetic_field_lower}) adopted in $r/R_\odot\le1.02$, we obtain the following condition.
\begin{align}
    B_{s,{\rm cb}} = 3 \left. B^{\rm PFSS} \right|_{r/R_\odot=1.02}.
\end{align}

Although a fraction of the magnetic field in the lower atmosphere should be highly inclined \citep{De_Pontieu_2004_Nature}, we simply assume that the flux tubes are vertical in the lower atmosphere.
In terms of $r$, the vertical magnetic field is represented by
\begin{align}
    r (s) = s,
\end{align}
for $1 \le r/R_\odot \le 1.02$.

\subsection{Basic equations and simulation setting \label{sec:basic_equations}}

Given the field-aligned profile of the background flux tube, the one-dimensional MHD equations with gravity, thermal conduction, and radiation are solved along the expanding flux tube.
The basic equations are given by \citep[see][for derivation]{Shoda_2021_arXiv}
\begin{align}
    &\frac{\partial \rho}{\partial t} + \frac{1}{r^2 f} \frac{\partial}{\partial s} \left( \rho v_s r^2 f \right) = 0, \\
    &\frac{\partial}{\partial t} \left( \rho v_s \right) + \frac{1}{r^2 f} \frac{\partial}{\partial s} \left[ \left( \rho v_s^2 + p + \frac{\vec{B}_\perp^2}{8 \pi} \right) r^2 f \right] \nonumber \\
    &= \left( p + \frac{1}{2} \rho \vec{v}_\perp^2 \right) \frac{d}{ds} \ln \left( r^2 f \right) - \rho \frac{GM_\odot}{r^2} \frac{dr}{ds}, \\
%\end{align}
%\begin{align}
    &\frac{\partial}{\partial t} \left( \rho \vec{v}_\perp \right) 
    + \frac{1}{r^2f} \frac{\partial}{\partial s} \left[ \left( \rho v_s \vec{v}_\perp - \frac{B_s \vec{B_\perp}}{4 \pi} \right) r^2 f \right] \nonumber \\
    &= \frac{1}{2} \left( - \rho v_s \vec{v}_\perp + \frac{B_s \vec{B}_\perp}{4 \pi} \right)
    \frac{d}{ds} \ln \left( r^2 f \right) + \rho \vec{D}^v_\perp, \\
    &\frac{\partial}{\partial t} \vec{B}_\perp + \frac{1}{r^2 f} \frac{\partial}{\partial s}\left[ \left( v_s \vec{B}_\perp - B_s \vec{v}_\perp \right) r^2 f \right] \nonumber \\
    &= \frac{1}{2} \left( v_s \vec{B}_\perp - B_s \vec{v}_\perp \right) \frac{d}{ds} \ln \left( r^2 f \right) + \sqrt{4 \pi \rho} \vec{D}^b_\perp, \\
    &\frac{\partial}{\partial t} e + \frac{1}{r^2f} \frac{\partial}{\partial s} \left[ \left( e + p_T \right) v_s - \frac{B_s}{4\pi} \left( \vec{v}_\perp \cdot \vec{B}_\perp \right) \right] \nonumber  \\
    &= - \rho v_s \frac{GM_\odot}{r^2} \frac{dr}{ds} + Q_{\rm cnd} + Q_{\rm rad},
\end{align}
where 
\begin{align}
    &\vec{v}_\perp = v_x \vec{e}_x + v_y \vec{e}_y, \hspace{2em}
    \vec{B}_\perp = B_x \vec{e}_x + B_y \vec{e}_y, \\
    &p_T = p + \frac{\vec{B}_\perp^2}{8\pi}, \hspace{2em} e = e_{\rm int} + \frac{1}{2} \rho \vec{v}^2 + \frac{\vec{B}_\perp^2}{8\pi}.
\end{align}
The term $\vec{D}_\perp^{v,b}$ denotes the phenomenological turbulent dissipation term \citep{Shoda_2018_ApJ_a_self-consistent_model}, which is given by
\begin{align}
    D^v_{x,y} &= - \frac{c_d}{4\lambda_\perp} \left( \left| z_{x,y}^+ \right| z_{x,y}^- + \left| z_{x,y}^- \right| z_{x,y}^+  \right), \label{eq:phenomenological_awt_vsource} \\ 
    D^b_{x,y} &= - \frac{c_d}{4\lambda_\perp} \left( \left| z_{x,y}^+ \right| z_{x,y}^- - \left| z_{x,y}^- \right| z_{x,y}^+  \right), \label{eq:phenomenological_awt_bsource}
\end{align}
where $z_{x,y}^\pm = v_{x,y} \mp B_{x,y}/\sqrt{4 \pi \rho}$.
The perpendicular correlation length is assumed to be proportional to the flux-tube radius.
\begin{align}
    \lambda_\perp = \lambda_{\perp,\odot} \frac{r}{R_\odot} \sqrt{\frac{f}{f_\odot}},
\end{align}
where we set the photospheric correlation length to 
$\lambda_{\perp,\odot} = 1,000 {\rm \ km}$.
$Q_{\rm cnd}$ and $Q_{\rm rad}$ represent the conduction heating per unit volume and radiative heating per unit volume, respectively.

Assuming that the gas is composed of hydrogen,
an equation of state with partially ionized hydrogen is used to connect the pressure, density, and temperature.
Considering the latent heat, the internal energy density is given by
\begin{align}
    e_{\rm int} = \frac{p}{\gamma-1} + \chi_{\rm H} n_{\rm H} I_{\rm H},   
\end{align}
where $\xi_{\rm H}$ is the ionization degree of hydrogen, $n_{\rm H}$ is the number density of hydrogen atom, and $I_{\rm H} = 13.6 {\rm \ eV}$ is the hydrogen ionization potential.
To obtain $\xi_{\rm H}$, 
we use an approximated Saha-Boltzmann equation.
The detailed procedure is described in \citet{Shoda_2021_arXiv}.

There are effects from partial ionization other than latent heat.
One of them is the enhanced dissipation of magnetic field, the ambipolar diffusion,
which is known to affect the propagation of high-frequency Alfv\'en waves in the chromosphere \citep{Khomenko_2014_PhPl}.
Such high-frequency waves are likely to be generated in a multi-dimensional system \citep[see discussion in][]{Cally_2018_ApJ}.
Because the system is one-dimensional in this work, 
the effect of partial ionization is not considered and remains to be seen.

The conductive heating is implemented in terms of conductive flux $q_{\rm cnd}$ by
\begin{align}
    Q_{\rm cnd} = - \frac{1}{r^2f} \frac{\partial}{\partial s} \left( q_{\rm cnd} r^2 f \right).
\end{align}
The Spitzer-H\"arm flux \citep{Spitzer_1953_PhysRev} is employed as the conductive flux.
To speed up the calculation without loss of reality, we quench the conductive flux in the distant region where $\rho$ is smaller than the critical value $\rho_{\rm sw}$.
The formulation of $q_{\rm cnd}$ is given by
\begin{align}
    q_{\rm cnd} = - \kappa_{\rm SH} T^{5/2} \frac{\partial T}{\partial s} \min \left( 1, \frac{\rho}{\rho_{\rm sw}} \right),
\end{align}
where $\kappa_{\rm SH} = 1.0 \times 10^{-6} {\rm \ erg \ cm \ K^{-7/2}}$.
We set $\rho_{\rm sw} = 10^{-20} {\rm \ g \ cm^{-3}}$.

The radiative heating is implemented as follows.
\begin{align}
    Q_{\rm rad} = - \frac{1}{\tau_{\rm rad}} \left( e_{\rm int} - e_{\rm int}^{\rm ref} \right) \left( 1 - \xi_{\rm rad} \right) + n_e n_{\rm H} \Lambda (T) \xi_{\rm rad},
\end{align}
where 
\begin{align}
    \tau_{\rm rad} = 0.1 {\rm \ s} \ \sqrt{\frac{\rho_\odot}{\rho}}, \ \ \ 
    \xi_{\rm rad} = \min \left[1, \exp \left( - 10\frac{p}{p_\odot} \right) \right],
\end{align}
where the subscript $\odot$ denotes the value at the photosphere.
The term $e_{\rm int}^{\rm ref}$ denotes the reference internal energy corresponding to the reference temperature $T^{\rm ref}$, and $\Lambda (T)$ is the optically-thin radiative loss function.
We set $T^{\rm ref} = T_\odot$.
To define $\Lambda(T)$ over a wide temperature range, we smoothly bridge the chromospheric \citep{Goodman_2012_ApJ} and coronal loss functions from the CHIANTI atomic database version 10 \citep{Dere_1997_AA,Del_Zanna_2021_ApJ} following \citet{Iijima_2016_PhD} (see \citet{Shoda_2021_arXiv} for details).

\subsection{Simulation domain and boundary condition}

The simulation domain extends from the photosphere and is aligned with the background magnetic field.
The radial distance of the outer boundary depends on the inclination of the magnetic field, but is always beyond $30R_\odot$.
The grid size $\Delta s$ is fixed to the minimum value $\Delta s_{\rm m}$ below the prescribed critical height $s_{\rm ge}$ and expands in $s$ until it reaches the maximum value $\Delta s_{\rm M}$.
Letting $s_i$ and $\Delta s_i$ be the position and size of the $i$-th grid, respectively,
we define $\Delta s_i$ as follows.
\begin{align}
    &\Delta s_i = \max \left[ \Delta s_{\rm m}, \min \left[ \Delta s_{\rm M}, \Delta s_{\rm m} + \frac{2\varepsilon_{\rm ge} \left( s_{i-1} - s_{\rm ge} \right)}{2 + \varepsilon_{\rm ge}} \right] \right], \nonumber \\
    &s_i = s_{i-1} + \frac{1}{2} \left( \Delta s_{i-1} + \Delta s_i \right), \ \ \ \ s_0 = R_\odot
\end{align}
where we set
\begin{align}
    &\Delta s_{\rm m} = 20 {\rm \ km}, \ \ \ \
    \Delta s_{\rm M} = 2000 {\rm \ km}, \nonumber \\
    &s_{\rm ge} = 1.04R_\odot, \ \ \ \ \ 
    \varepsilon_{\rm ge} = 0.01. \nonumber
\end{align}

At the outer boundary, because the solar wind becomes supersonic and super-Alfv\'enic,
any MHD waves propagate only in the outward direction.
For this reason, we impose the free boundary conditions as
\begin{align}
    \left. \frac{\partial \rho}{\partial s} \right|_{\rm out} = 0, \ \ \ 
    \left. \frac{\partial \vec{v}}{\partial s} \right|_{\rm out} = 0, \ \ \ 
    \left. \frac{\partial \vec{B}_\perp}{\partial  s}\right|_{\rm out} = 0,
\end{align}
where the subscript ``out" denotes the value at the outer boundary.
Because heat can propagate backward against the supersonic flow by thermal conduction, it would be risky to impose a free boundary condition for $p$ or $e_{\rm int}$.
Instead, we impose the boundary condition for $e_{\rm int}$ as follows:
\begin{align}1
    e_{\rm int} \propto r^{-1/2} \ \ \ \left( r \ge r_{\rm out} \right).
\end{align}

At the inner boundary, the temperature is fixed to the solar effective temperature.
\begin{align}
    T_\odot = 5.77 \times 10^3 {\rm \ K}, 
\end{align}
where the subscript $\odot$ denotes the value at the inner boundary.
Some solar observations reveal that the photospheric magnetic fields are localized in kilo-Gauss patches \citep{Keller_2004_ApJ,Tsuneta_2008_ApJ}.
We therefore fix the photospheric field strength to 
\begin{align}
    B_{s,\odot} = 1.34 \times 10^3 {\rm \ G}.
\end{align}

To excite upward longitudinal waves at the bottom boundary, 
we impose the time-dependent boundary conditions for $\rho$ and $v_s$ as follows:
\begin{align}
    \rho_\odot = \overline{\rho_\odot} \left( 1 + \frac{\delta v_\odot}{a_\odot} \right), \ \ \ 
    v_{s,\odot} = \delta v_{s,\odot},
\end{align}
where $\delta v_{s,\odot}$ denotes the longitudinal velocity fluctuation at the photosphere.
$\overline{\rho_\odot}$ represents the time-averaged mass density, and $a_\odot$ is the sound speed in the photosphere, which are given, respectively, by
\begin{align}
    \overline{\rho_\odot} &= 1.88 \times 10^{-7} {\rm \ g \ cm^{-3}}, \ \ \ 
    a_\odot = 8.91 {\rm \ km \ s^{-1}}.
\end{align}
The transverse fluctuations are provided in terms of Els\"asser variables defined by
\begin{align}
    \vec{z}_\perp^\pm = \vec{v}_\perp \mp \frac{\vec{B}_\perp}{\sqrt{4 \pi \rho}}.
\end{align}
The free boundary condition is imposed on the inward Els\"asser variables
\begin{align}
    \left. \frac{\partial \vec{z}_\perp^{-}}{\partial s} \right|_\odot = 0.
\end{align}
For the upward Els\"asser variables, the time-dependent boundary condition is imposed to excite Alfv\'en waves at the photosphere.
\begin{align}
    z^+_{x,\odot} = 2 \delta v_{x,\odot}, \ \ \ z^+_{y,\odot} = 2 \delta v_{y,\odot}
\end{align}

\begin{figure*}[t!]
\centering
\hspace{1em} \includegraphics[width=150mm]{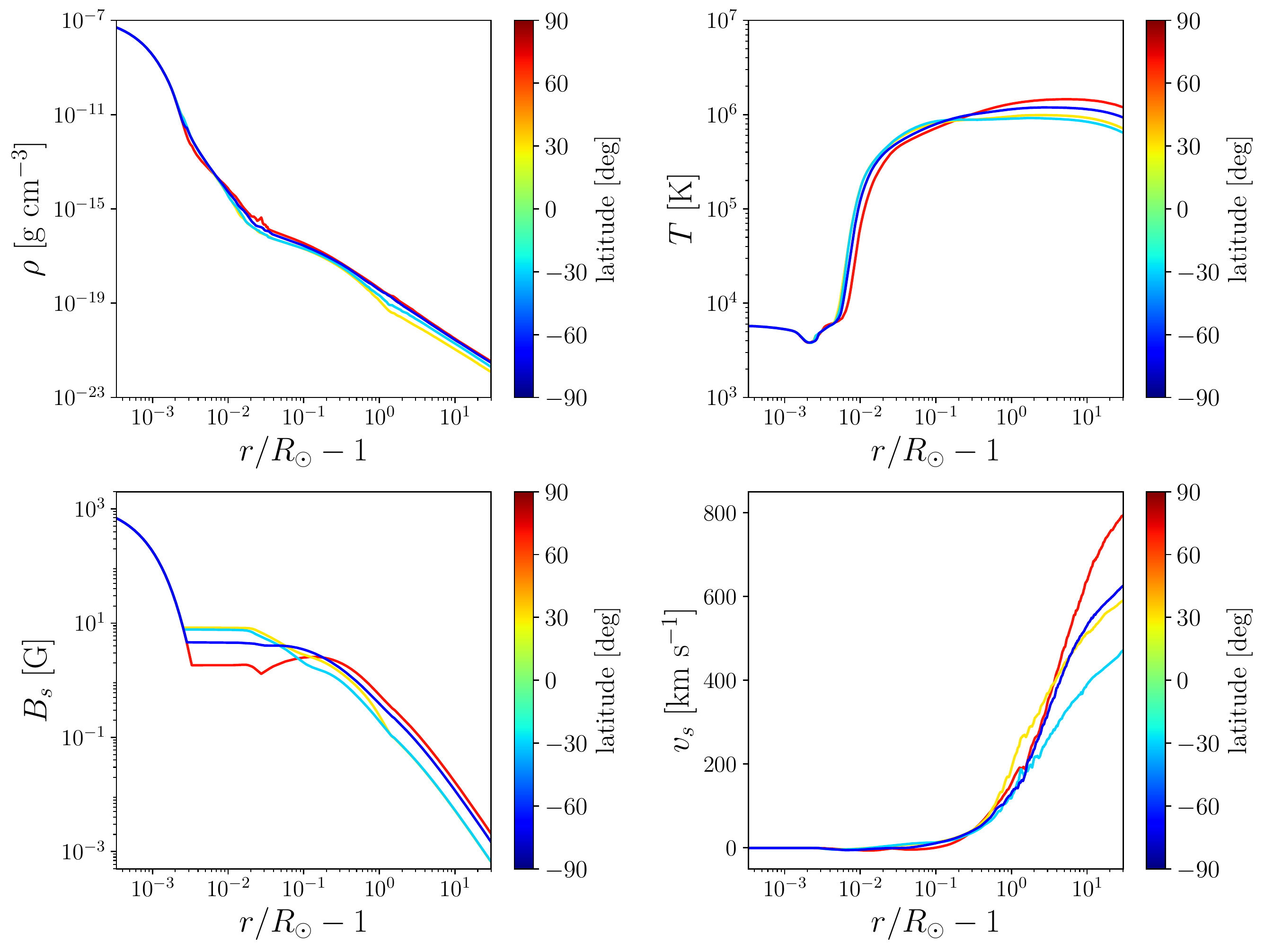}
\vspace{0em}
\caption{
Time-averaged radial structures of the simulated solar wind for CR 2218 (minimal activity phase).
The four panels show the mass density ($\rho$, left top), temperature ($T$, right top), background magnetic field ($B_s$, left bottom), and field-aligned velocity ($v_s$, right bottom), respectively.
The line colors indicate the corresponding latitudes ($70^\circ, 30^\circ, -30^\circ, -70^\circ$) at the source surface.
The longitude is fixed to $300^\circ$.
}
\label{fig:radial_profile_comparison}
\vspace{2em}
\end{figure*}

The velocity fluctuations are given in terms of broadband fluctuation by
\begin{align}
    \delta v_{s,x,y,\odot} &\propto \sum_{i=0}^N \sin \left( 2 \pi f^i t + \phi^i_{s,x,y}  \right) / \sqrt{f^i}, \\
    f^i &= \frac{\left( N-i\right) f_{\rm min} + i f_{\rm max}}{N},
\end{align}
where $\phi^i_{s,x,y}$ represents a random phase function, $N+1$ is the total number of modes, and
\begin{align}
    f_{\rm min} = 1.00 \times 10^{-3} {\rm \ Hz}, \ \ \ f_{\rm min} = 1.00 \times 10^{-2} {\rm \ Hz},
\end{align}
are the minimum and maximum wave frequencies.
In this study, we set $N=100$.
The root-mean-squared amplitudes of $\delta v_{s,x,y,\odot}$ were tuned to $0.4 {\rm \ km \ s^{-1}}$, that is,
\begin{align}
    \sqrt{\overline{\delta v_{s,x,y,\odot}^2}} = 0.4 {\rm \ km \ s^{-1}},
\end{align}
where the overline denotes the time averaging.

\section{Result}

\begin{figure*}[t!]
\centering
\includegraphics[width=170mm]{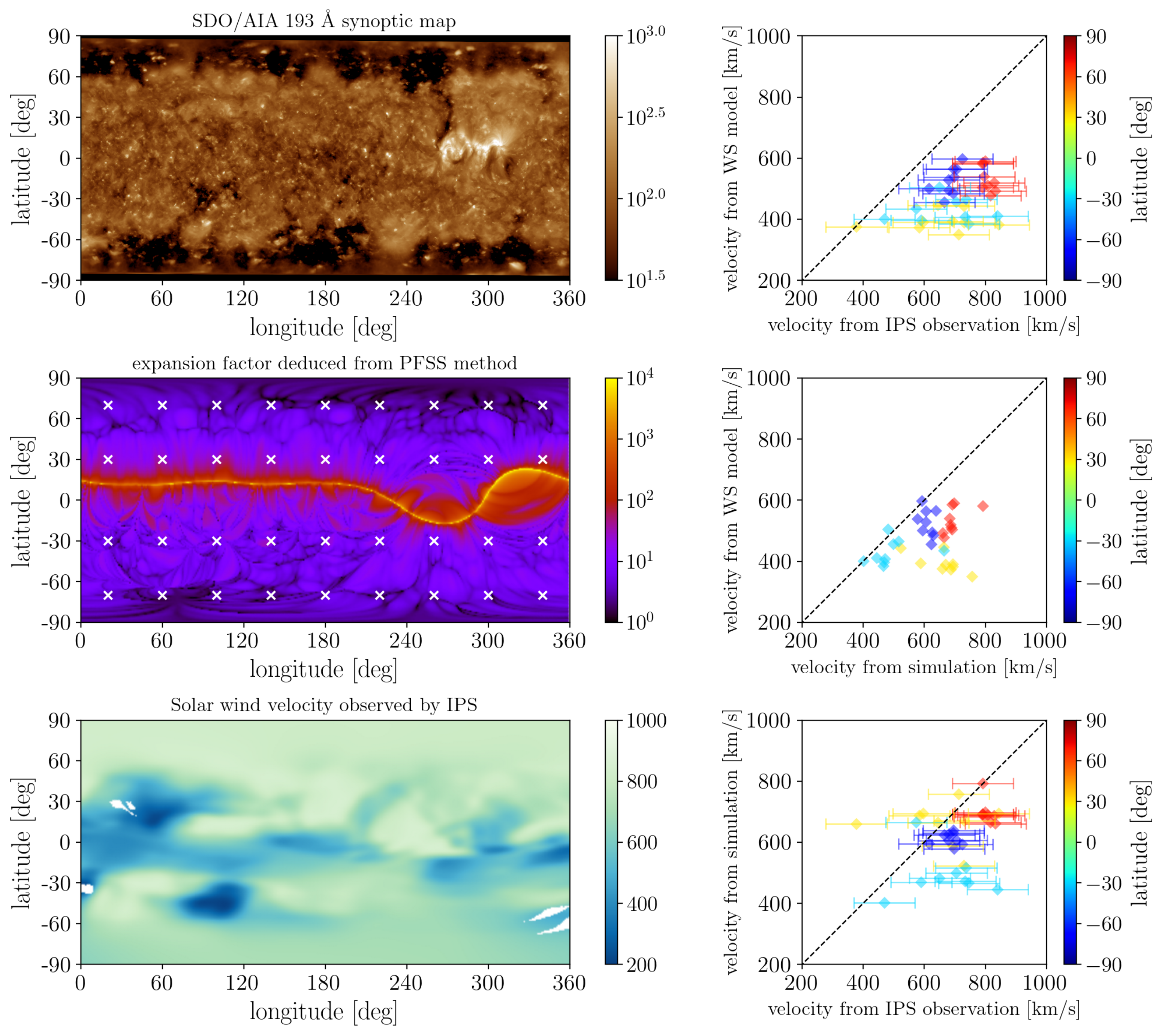}
\vspace{0em}
\caption{
Overview of CR 2218 results.
Left-top: synoptic map constructed from SDO/AIA 193 ${\rm \AA}$ observation.
Left-middle: flux-tube expansion factor at the source surface ($r = r_{\rm SS}$) deduced from the PFSS extrapolation.
We use $r_{\rm SS}/R_\odot = 2.5$.
Left-bottom: solar wind velocity measured by IPS technique.
Right panels: scattered plots showing the correlations between $v_{\rm SIM}$, $v_{\rm WS}$, and $v_{\rm IPS}$.
One symbol corresponds to one simulation run, and the color represents the latitude of the wind.
The error in $v_{\rm IPS}$ was fixed at $100 {\rm \ km \ s^{-1}}$.
}
\label{fig:run_selection2_CR2218}
\vspace{2em}
\end{figure*}

\subsection{Radial structure of the simulated solar wind}

Clarifying the detailed physical mechanism of the solar wind heating and acceleration is not the main scope of this study \citep[see, for instance,][for a detailed discussion of the physical mechanism]{Shoda_2018_ApJ_a_self-consistent_model,Shoda_2019_ApJ}.
Rather than detailed analyses, we show some examples of the radial structure of the simulated solar wind to grasp an overview of the simulation results.

Figure \ref{fig:radial_profile_comparison} shows the time-averaged radial profiles of the simulated solar wind for Carrington rotation (CR) 2218 in which the solar activity was close to the minimum phase.
The four lines correspond to $70^\circ$, $30^\circ$, $-30^\circ$, and $-70^\circ$ in terms of latitude, as indicated by the line color.
The longitude is fixed to $300^\circ$.
Higher-/lower-speed winds are found in higher/lower latitudes, 
which is consistent with the observed latitudinal structure in the solar minimum  \citep{McComas_2003_GRL}.

Observing closely the temperature profiles (right-top panel), noticeable differences between fast and slow winds are present.
Near the base of the corona ($r/R_\odot-1 \le 0.1$), 
slower winds (yellow and cyan) were hotter than faster winds (red and blue).
This relationship is reversed beyond the wind acceleration region ($r/R_\odot-1 \ge 1$).
This behavior is consistent with the fact that the fast solar wind exhibits a larger in-situ proton temperature and smaller freezing-in (coronal base) electron temperature than the slow solar wind
\citep[see][and references there in]{Cranmer_2017_SSRev}.

The relation between the wind velocity and magnetic field is shown in the bottom two panels of Figure \ref{fig:radial_profile_comparison}.
The field strength of the faster wind is smaller near the coronal base ($r/R_\odot-1 \le 0.1$), but larger in the solar wind ($r/R_\odot-1 \ge 1$).
In terms of the expansion factor, the simulation results imply that a small expansion of flux tube yields a fast solar wind. 
This relation is consistent with the classical observational trend \citep{Wang_1990_ApJ,Arge_2000_JGR},
although the role of expansion factor is recently questioned \citep{Riley_2015_SpaceWeather}.

In brief, our model reproduces 
1. the latitudinal distribution of the wind speed,
2. the relationship between the coronal/wind temperature and wind speed, and
3. the inverse correlation between the expansion factor and wind speed.
These agreements validate the WTD model in that it accounts for the observed qualitative behavior of the solar wind.
To more quantitatively test the model, in the following sections,
we compare the simulation results with the IPS measurements. 

\begin{figure*}[t!]
\centering
\includegraphics[width=170mm]{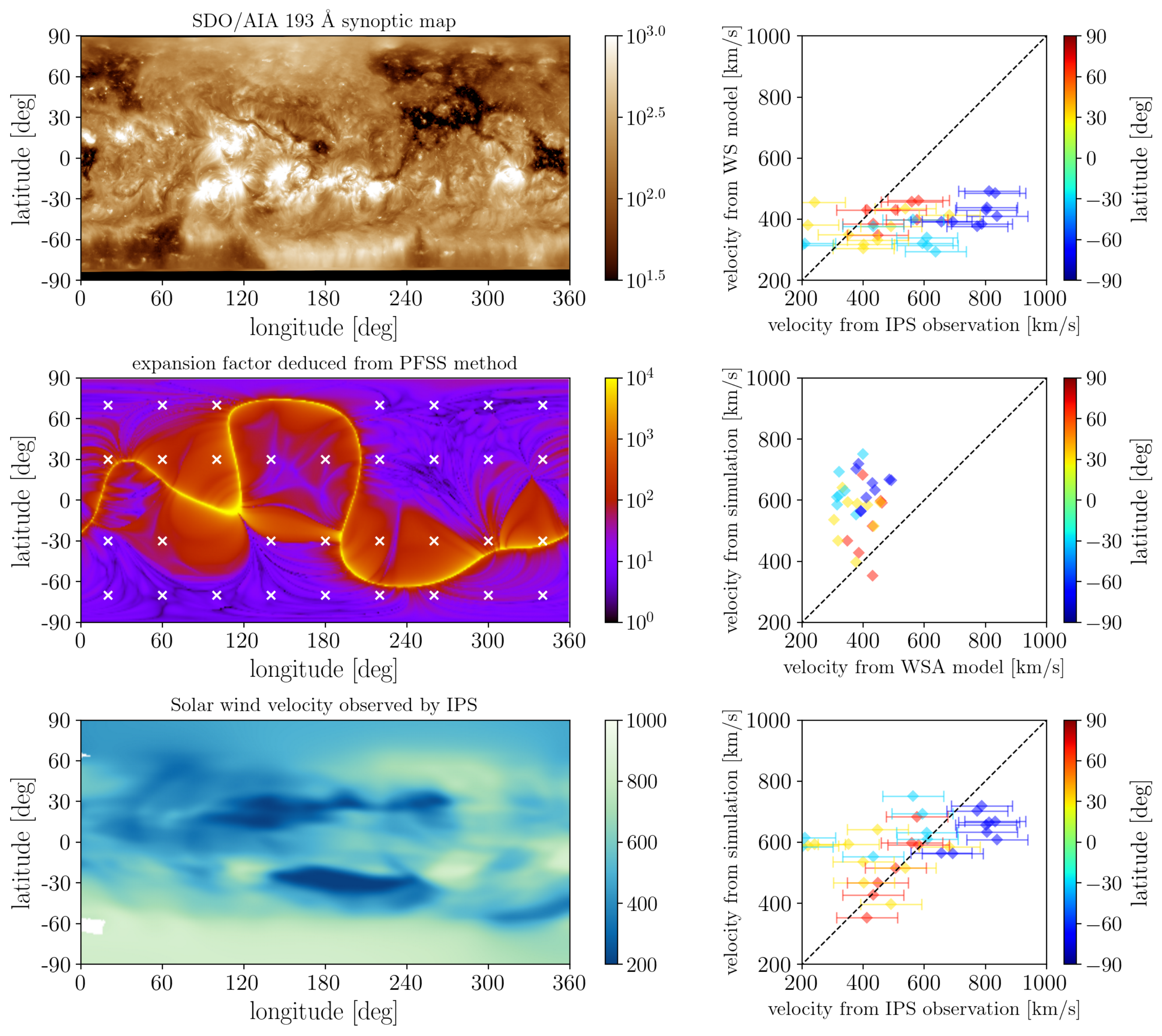}
\vspace{0em}
\caption{
Same as Figure \ref{fig:run_selection2_CR2218} but for CR 2169.
}
\label{fig:run_selection2_CR2169}
\vspace{2em}
\end{figure*}

\subsection{Solar wind observation with the IPS}
The IPS is a radio-scattering phenomenon caused by turbulence in the solar wind. 
We used the IPS data taken by the Institute for Space-Earth Environmental Research (ISEE), Nagoya University \citep{Tokumaru_2011_RadioScience}, 
which is composed of three radio telescopes at 327 MHz.
The solar wind speed is derived by the cross-correlations of the IPS data between the three stations. 
The global distribution of the solar wind is derived by projecting the solar wind speed data onto the source surface using the tomography method \citep{Jackson_1998_JGR,Kojima_1998_JGR}, 
which has been recently improved by \citet{Tokumaru_2021_ApJ}. 
The tomography method excludes transient phenomena such as coronal mass ejections that are also detected by IPS observations \citep[for instance,][]{Iwai_2019_EPS,Iwai_2021_EPS}. 
Therefore, the tomography method derives the solar wind velocity of the background solar wind at a steady state.

The root-mean-squared discrepancy between the IPS and the in-situ measurement is typically 100--150 ${\rm km \ s^{-1}}$.
The spatial resolution of the IPS data on the synoptic map is approximately 15$^\circ$. 
Hence, we averaged the derived IPS data within the 15$^\circ$ $\times$ 15$^\circ$ region and recognized it as a typical solar wind speed. 
Note that this low resolution of the solar wind map could create an artificial middle-speed wind around the boundary between the fast and slow wind regions, 
even if the solar wind would be a bi-modal fast and slow wind structure. 
Therefore, the solar wind speed close to the HCS may have a larger error.

\subsection{Solar wind in an inactive phase (CR 2218) \label{sec:result:CR2218}}

Carrington rotation (CR) 2218 is half a year before the end of the solar cycle 24 in which the monthly averaged relative sunspot number is 1.2,
and behaves as a typical phase of the minimal activity.
Because the IPS observation and its tomographic analysis provide the nearly full solar wind velocity distribution on the Carrington map, we focus on CR 2218 as an example of a magnetically inactive phase.

Left panels in Figure \ref{fig:run_selection2_CR2218} show 
the synoptic maps of the SDO/AIA 193 image (top), 
the expansion factor $f_{\rm exp}$ deduced from the PFSS method, 
and the solar wind velocity measured by the IPS observation projected on the source surface ($r/R_\odot=2.5$).
The AIA synoptic image is made from the open dataset by \citet{Hamada_2020_SolPhys}.
The simulations were performed for the selected points indicated by the white crosses in the $f_{\rm exp}$ maps.
A low level of magnetic activity is observed from the large coronal holes developed at high latitudes (observed in the AIA image) 
and the current sheet (indicated by the large-$f_{\rm exp}$ region) remaining at low latitudes.
As a typical structure of the solar wind in the activity-minimum, high- and low-latitude regions are dominated by fast and slow solar winds, respectively (as seen in the IPS map).

The three right in Figure \ref{fig:run_selection2_CR2218} show the correlations between the three solar wind velocities: $v_{\rm SIM}$, $v_{\rm WS}$, and $v_{\rm IPS}$.
$v_{\rm SIM}$ denotes the wind speed from the simulation, which is defined by the time-averaged field-aligned velocity at $r/R_\odot=30$, that is,
\begin{align}
    v_{\rm SIM} =  \left. \overline{v_s} \right|_{r/R_\odot=30},
\end{align}
where the overline denotes the time averaging.
$v_{\rm WS}$ is the wind speed predicted by the empirical Wang--Sheeley model \citep{Wang_1990_ApJ}.
In this study, we use the relation by \citet{Arge_2000_JGR}, that is,
\begin{align}
    v_{\rm WS} = 267.5 + \frac{410.0}{{f_{\rm exp}}^{2/5}} {\rm \ \ km \ s^{-1}},
\end{align}
where $f_{\rm exp}$ denotes the expansion factor.
$v_{\rm IPS}$ denotes the IPS measurement of the wind speed.

The three scattered plots show the following results.
\begin{enumerate}
    \item The Wang--Sheeley empirical model ($v_{\rm WS}$) is systematically lower than the observed velocity ($v_{\rm IPS}$).
    The deviation between the two is larger for high-latitude cases.
    \item The simulated wind velocity ($v_{\rm SIM}$) exhibits a better agreement with the IPS measurement ($v_{\rm IPS}$) than with the Wang--Sheeley empirical model ($v_{\rm WS}$).
    The agreement is particularly good in high-latitude cases.
\end{enumerate}
The Wang--Sheeley model shows better performance in the lower-latitude region, and it is a natural consequence because the model is calibrated by near-Earth observations.
From these scattered plots, our model was found to reproduce the observed wind velocity with a better accuracy than the Wang--Sheeley model, at least in high-latitude regions.
We investigate whether this trend is also found in the activity-maximum phase in the following section.

\begin{table*}[t!]
\centering
  % defining horizontal cell size
  \begin{tabular}{p{12.0em} p{12.0em} p{9.0em} p{9.0em} p{9.0em}}
    \begin{tabular}{l} \hspace{-3.0em} Carrington Rotation \end{tabular} 
    & \begin{tabular}{l} \hspace{-3.0em} velcoity model \end{tabular}
    & \begin{tabular}{c} $\Delta v_{\rm rms} \ {\rm [km \ s^{-1}]}$ \hspace{-1.5em} \\ (high-lat cases) \hspace{-1.5em} \end{tabular}
    & \begin{tabular}{c} $\Delta v_{\rm rms} \ {\rm [km \ s^{-1}]}$ \hspace{-1.5em} \\ (mid-lat cases) \hspace{-1.5em}
    \end{tabular}
    & \begin{tabular}{c} $\Delta v_{\rm rms} \ {\rm [km \ s^{-1}]}$ \hspace{-1.5em} \\ (all cases) \hspace{-1.5em} \end{tabular}
    \rule[-5.5pt]{0pt}{20pt} \\ \hline \hline
    % 1st row, CR 2169, rss=2.5Rs, ips-sim comparison
    \hspace{-0.5em} CR 2169 (active phase)
    & simulation
    & \hspace{4.0em} $112.8$ 
    & \hspace{4.0em} $171.4$ 
    & \hspace{4.0em} $135.2$ \rule[-5.5pt]{0pt}{20pt} \\
    % 2nd row, CR 2169, rss=2.5Rs, ips-wsa comparison
    \hspace{-0.5em} CR 2169 (active phase)
    & Wang-Sheeley relation
    & \hspace{4.0em} $277.5$ 
    & \hspace{4.0em} $156.1$ 
    & \hspace{4.0em} $243.8$ \rule[-5.5pt]{0pt}{20pt} \\
    % 3rd row, CR 2218, rss=2.5Rs, ips-sim comparison
    \hspace{-0.5em} CR 2218 (inactive phase)
    & simulation
    & \hspace{4.0em} $103.0$ 
    & \hspace{4.0em} $185.0$ 
    & \hspace{4.0em} $149.7$ \rule[-5.5pt]{0pt}{20pt} \\
    % 4th row, CR 2218, rss=2.5Rs, ips-wsa comparison
    \hspace{-0.5em} CR 2218 (inactive phase)
    & Wang-Sheeley relation
    & \hspace{4.0em} $227.7$ 
    & \hspace{4.0em} $278.4$ 
    & \hspace{4.0em} $254.3$ \rule[-5.5pt]{0pt}{20pt} \\
  \end{tabular}
  \vspace{1.0em}
  \caption{This table summarizes the root-mean-squared errors of the solar wind velocity $\Delta v_{\rm rms}$ for the simulation and Wang-Sheeley relation.
  The value of $\Delta v_{\rm rms}$ is calculated for high-latitude cases (latitude $=\pm 70^\circ$), mid-latitude cases (latitude $=\pm 30^\circ$), and all cases (latitude $=\pm 30^\circ, \ \pm 70^\circ$).}
  \vspace{1.0em}
  \label{table:dv_rms}
\end{table*}

\subsection{Solar wind in an active phase (CR 2169) \label{sec:result:CR2169}}

Carrington Rotation (CR) 2169, beginning on October 4th and ending on October 31st in 2015, is 1.5 years after the sunspot-number maximum in solar cycle 24 with the monthly averaged relative sunspot number of 63.6.
Because the global structure of the solar wind is obtained by the IPS during this rotation,
we focus on CR 2169 as an example of a magnetically active phase.

Figure \ref{fig:run_selection2_CR2169} shows the observation and simulation results in the same format as Figure \ref{fig:run_selection2_CR2218} but for CR 2169.
Unlike CR 2218, a high level of magnetic activity is observed in the AIA coronal image with several bright structures at low latitudes.
As seen in the IPS map, a fraction of the northern polar region is dominated by slow solar wind, which is also typical of the magnetically active phase.

As shown in Figure \ref{fig:run_selection2_CR2218}, the simulation points are indicated by the white crosses in the $f_{\rm exp}$ map.
Simulations at $({\rm longitude, \ latitude}) = (100^\circ, \ -30^\circ)$, $(140^\circ, \ 70^\circ)$, and $(180^\circ, \ 70^\circ)$ are not performed because, in these cases, the PFSS extrapolation yields unrealistic discontinuities in the field strength near the source surface.

The scatter plots on the right-hand side of Figure \ref{fig:run_selection2_CR2169} exhibit a similar trend to that shown in \ref{fig:run_selection2_CR2218}.
The Wang--Sheeley relation tends to predict slower wind speeds than the IPS measurement, 
while the agreement is much better between the simulation and IPS, particularly in the high-latitude region.
In contrast to CR 2218, in which both polar regions are dominated by fast solar wind,
a fraction of the northern polar region is covered by the slow solar wind.
Interestingly, the slow solar wind in the northern polar region is reproduced by our simulation, as shown by the red symbols in the right-bottom panel of Figure \ref{fig:run_selection2_CR2169}.

In the mid-latitude cases, the Wang--Sheeley relation and simulation exhibited opposite trends with respect to the IPS measurement.
The Wang--Sheeley relation tends to underestimate the wind velocity,
whereas the simulation tends to overestimate it.
The two methods yield a similar level of root-mean-squared deviation from the IPS measurements
As in CR 2218, we find that the Wang--Sheeley relation performs better in the lower-latitude regions.

\subsection{Performance in the wind velocity prediction}

To evaluate the performance of the Wang-Sheeley relation and simulation in the wind velocity prediction, 
we calculated the root-mean-squared error $\Delta v_{\rm rms}$ with respect to the IPS measurement as follows.
\begin{align}
    \Delta v_{\rm rms} = \sqrt{\frac{1}{N} \sum_{i=1}^{N} \left( v_{\rm WS, \ SIM}^i - v_{\rm IPS}^i \right)^2},
\end{align}
where $N$ denotes the total number of simulation runs and the superscript $i$ denotes the $i$-th simulation run.
$\Delta v_{\rm rms}$ is calculated separately for high-latitude, mid-latitude, and all runs for each CR.
The results of the error calculation are summarized in Table \ref{table:dv_rms}.

In high-latitude cases, the root-mean-squared error of the simulation is $\Delta v_{\rm rms} \approx 100 {\rm \ km \ s^{-1}}$, approximately in the same magnitude as the error in the IPS observations.
Interestingly, CR 2169 and 2218 yielded nearly the same value of the simulation error despite the different levels of magnetic activity.
The Wang--Sheeley relation yields a much larger error, 
typically $\Delta v_{\rm rms} \approx 250 {\rm \ km \ s^{-1}}$,
with a larger error found in the active phase (CR 2169).

In mid-latitude cases, the accuracy of the simulation decreases,
as already discussed in Sections \ref{sec:result:CR2218} and \ref{sec:result:CR2169}.
In CR 2169, the Wang--Sheeley relation exhibits better performance than the simulation.
In contrast, in CR 2218, the simulation is still more accurate than the Wang--Sheeley relation.

The all-case root-mean-squared errors are $\Delta v_{\rm rms} \approx 140 {\rm \ km \ s^{-1}}$ for the simulation,
and $\Delta v_{\rm rms} \approx 250 {\rm \ km \ s^{-1}}$ for the Wang--Sheeley relation.
From the data provided here, 
it would be risky to conclude that the simulation is more capable of predicting the wind velocity than the Wang-Sheeley relation.
To discuss the performance of the solar wind prediction, we need to consider a large number of CRs with a range of magnetic activity.
The Wang--Sheeley relation should also be calibrated to maximize its performance, because its best-fit model varies over time \citep{Riley_2015_SpaceWeather}.
Instead, we conclude that the Alfv\'en-wave model of the solar wind has the potential to predict the solar wind velocity in a first-principle manner, at least in high-latitude regions.

\section{Summary and Discussion}

In this study, we test the theoretical (wave/turbulence-driven, WTD) model of the solar wind using IPS observations.
Along background field lines inferred by the PFSS method, we directly solved the MHD equations from the photosphere to the solar wind.
The simulated solar wind velocity was directly compared with the IPS measurements to validate the model performance.

Regardless of the activity level, the WTD model of the solar wind accurately reproduces the observed wind velocity at high latitudes, with a typical root-mean-squared error of $100 {\rm \ km \ s^{-1}}$.
Given that the typical error in the IPS measurement is $100-150 {\rm \ km \ s^{-1}}$,
we conclude that the WTD model captures the structure of high-latitude solar wind.
In terms of the acceleration mechanism, our results indicate that the high-latitude solar wind is accelerated by Alfv\'en waves, even in the activity maximum phase.
Indeed, the slow solar wind observed in the northern polar region of CR 2169 was well reproduced by the model.

In the mid-latitude cases, the WTD model becomes less accurate, with a typical root-mean-squared error of $180 {\rm \ km \ s^{-1}}$.
There are three possible reasons for this reduction in accuracy.

First, physical mechanisms other than Alfv\'en-wave heating and acceleration work in the lower-latitude regions.
One promising mechanism for the additional energy release is interchange reconnection \citep{Fisk_2003_JGR,Antiochos_2011_ApJ}.
Indeed, some solar observations indicate the presence of localized outflows near the edge of active regions \citep{Sakao_2007_Science,Harra_2008_ApJ,Brooks_2011_ApJ,Brooks_2020_ApJ,Brooks_2021_ApJ},
which are found to feed a non-negligible fraction of mass flux into the solar wind \citep{Brooks_2015_NatreCommunications}.
To improve the model accuracy,
reconnection-driven energy release must be considered \citep{Lionello_2016_ApJ,Cranmer_2018_ApJ} as well as granulation-driven Alfv\'en waves \citep{Wang_2020_ApJ}.

Second, magnetic-field extrapolation could be inaccurate in the lower-latitude region.
Although the PFSS method is found to reproduce a large-scale magnetic field without a large numerical cost \citep{Riley_2006_ApJ},
it is no longer validated near the current sheets that tend to appear in mid- to low-latitude regions.
A possible improvement is to adopt the Schatten current sheet model \citep[SCS,][]{Schatten_1971}, 
which is found to account for the latitudinal dependence of the interplanetary magnetic field beyond the PFSS model \citep{Wang_1995_ApJ}.
We note that, when connecting the PFSS and SCS models, careful treatment is required to reduce the discontinuity in the magnetic field \citep{McGregor_2008_JGR,Reiss_2020_ApJ}.
Another factor that affects the accuracy of magnetic field extrapolation is the value of the source surface radius $r_{\rm SS}$.
Although we adopt the conventional value of $r_{\rm SS}$ \citep[$r_{\rm SS}/R_\odot = 2.5$,][]{Hoeksema_1983_JGR},
a recent {\it Parker Solar Probe} observation reveals that the source surface can be much closer \citep{Badman_2020_ApJ}.
The photospheric magnetogram can also be improved, from GONG observations to the data-assimilated flux-transport model \citep[ADAPT,][]{Arge_2010_AIPC}.
In fact, the open-flux problem could be (partly) solved by adding magnetic flux in the polar region \citep{Riley_2019_ApJ}, 
which indicates that the observed magnetic field in the polar region is still inaccurate.
These uncertainties need to be discussed quantitatively in future.
 
Finally, there is also a possibility that the spatial resolution of the IPS observation was not sufficient to resolve the complicated solar wind source region in the active phase case. 
The improved IPS observation could give a better spatial resolution of the tomographic map in future.

In spite of the limitations mentioned above,
the high-latitude solar wind is well reproduced by the physics-based WTD model.
In addition to the novel validation of the WTD model, 
this study opens the possibility of a physics-based photosphere-wind connection beyond the current transition region-wind connection \citep{van_der_Holst_2014_ApJ,Usmanov_2018_ApJ}.
Most of the current space-weather models employ the empirical solar wind velocity \citep{Riley_2001_JGR,Riley_2015_SpaceWeather} at the inner boundary. \citep{Odstrcil_2003_AdSpR,Shiota_2014_SpaceWeather,Cash_2015_SpaceWeather,Wold_2018_JSWSC}.
Improvements of this work could lead to 
the theoretical validation of the empirical models \citep[e.g. WSA model,][]{Arge_2004_JASTP,Riley_2015_SpaceWeather}, and ideally,
improved forecasting of space weather.

Numerical computations were carried out on the Cray XC50 at the Center for Computational Astrophysics (CfCA), National Astronomical Observatory of Japan.
IPS observations were performed under the solar wind program of the Institute for Space-Earth Environmental Research, Nagoya University.
M.S. is supported by a Grant-in-Aid for Japan Society for the Promotion of Science (JSPS) Fellows and by the NINS program for cross-disciplinary study (grant Nos. 01321802 and 01311904) on Turbulence, Transport, and Heating Dynamics in Laboratory and Solar/ Astrophysical Plasmas: “SoLaBo-X.”
K.I. is supported by JSPS KAKENHI grant Nos. 19K22028 and 21H04517.
D.S. is supported by JSPS KAKENHI grant Nos. 19K23472 and 21H04492.
This work made use of matplotlib, a Python library for publication quality graphics \citep{Hunter_2007_CSE}, and NumPy \citep{van_der_Walt_2011_CSE}.

\begin{appendix}
\vspace{-1em}
\section{Uncertainty in the chromospheric magnetic field} \label{app:chrom_mag}
In all the simulation runs displayed above, 
we have assumed that the magnetic field expands to keep the plasma beta nearly unity from the photosphere to the chromosphere (see Eq~\eqref{eq:magetic_field_lower}).
This assumption is validated by neither simulation nor observation, and thus, 
we should discuss how the solar wind speed is affected by the choice of the magnetic-field profile in the lower atmosphere.
For this purpose, we have performed a parameter survey on the chromospheric magnetic field for a fixed latitude, longitude, and Carrington rotation.
Specifically, we focus on $({\rm longitude}, {\rm latitude}) = (20^\circ, 20^\circ)$ in CR 2218.

As in Eq~\eqref{eq:magetic_field_lower},
we assume that the magnetic field exponentially decreases as a function of height until it reaches the basal (coronal) value. 
\begin{align}
    B_s \left( s \right) = \max \left[ B_{s,{\rm cb}}, B_{s,0} \exp 
    \left( - \frac{r -R_\odot}{c_{\rm mag}H_\odot}  \right) \right].
    \label{eq:mag_appendix}
\end{align}
We note that, letting $B_{s,0}=B_{s,\odot}=1340{\rm \ G}$ and $c_{\rm mag} =2.0$,
Eq.~\eqref{eq:mag_appendix} is equivalent to Eq~\eqref{eq:magetic_field_lower}.
We have performed a series of simulation with respect to $B_{s,0}$ and $c_{\rm mag}$, 
the list of which is shown in Table \ref{table:appendix_list}.

\begin{table}[t!]
\centering
  % defining horizontal cell size
  \begin{tabular}{p{6em} p{5em} p{6em} p{7em}}
    $B_{s,0} {\rm \ [G]}$
    & $c_{\rm mag}$
    & \hspace{-1em} $v_r {\rm \ [km \ s^{-1}]}$
    & \hspace{0em} $\dot{M}_w  \ [M_\odot {\rm \ yr^{-1}]}$
    \rule[-5.5pt]{0pt}{10pt} \\ \hline \hline
    % fiducial case, B_{s,0}=1340 G, c_mag=2.0
    \hspace{-0.25em} 1340
    & \hspace{0.0em} 2.0
    & \hspace{0.5em} $692$ 
    & \hspace{0.5em} $1.78 \times 10^{-14}$ 
    \rule[-5.5pt]{0pt}{20pt} \\
    % fast expansion, B_{s,0}=1340 G, c_mag=1.5
    \hspace{-0.25em} 1340
    & \hspace{0.0em} 1.5
    & \hspace{0.5em} $658$ 
    & \hspace{0.5em} $1.51 \times 10^{-14}$
    \rule[-5.5pt]{0pt}{20pt} \\
    % slow expansion, B_{s,0}=1340 G, c_mag=2.5
    \hspace{-0.25em} 1340
    & \hspace{0.0em} 2.5
    & \hspace{0.5em} $745$ 
    & \hspace{0.5em} $1.49 \times 10^{-14}$
    \rule[-5.5pt]{0pt}{20pt} \\
    % weak_hmag2.0, B_{s,0}=335 G, c_mag=2.0
    \hspace{-0.0em} 335
    & \hspace{0.0em} 2.0
    & \hspace{0.5em} $611$ 
    & \hspace{0.5em} $2.01 \times 10^{-14}$
    \rule[-5.5pt]{0pt}{20pt} \\
    % weak_hmag3.0, B_{s,0}=335 G, c_mag=3.0
    \hspace{-0.0em} 335
    & \hspace{0.0em} 3.0
    & \hspace{0.5em} $620$ 
    & \hspace{0.5em} $2.38 \times 10^{-14}$
    \rule[-5.5pt]{0pt}{20pt} \\
  \end{tabular}
  \vspace{1.0em}
  \caption{List of the simulation runs in this survey.
  The wind velocity ($v_r$) and the mass-loss rate ($\dot{M}_w = 4 \pi r^2 \rho v_r$) are measured at $r/R_\odot =30$ and time averaged.
  %We note that the input parameters of this simulation are $B_{s,0}$ and $c_{\rm mag}$ and the output parameters are $v_r$ and $\dot{M}_w$.
  }
  \vspace{1.0em}
  \label{table:appendix_list}
\end{table}

\begin{figure*}[t!]
\centering
\includegraphics[width=180mm]{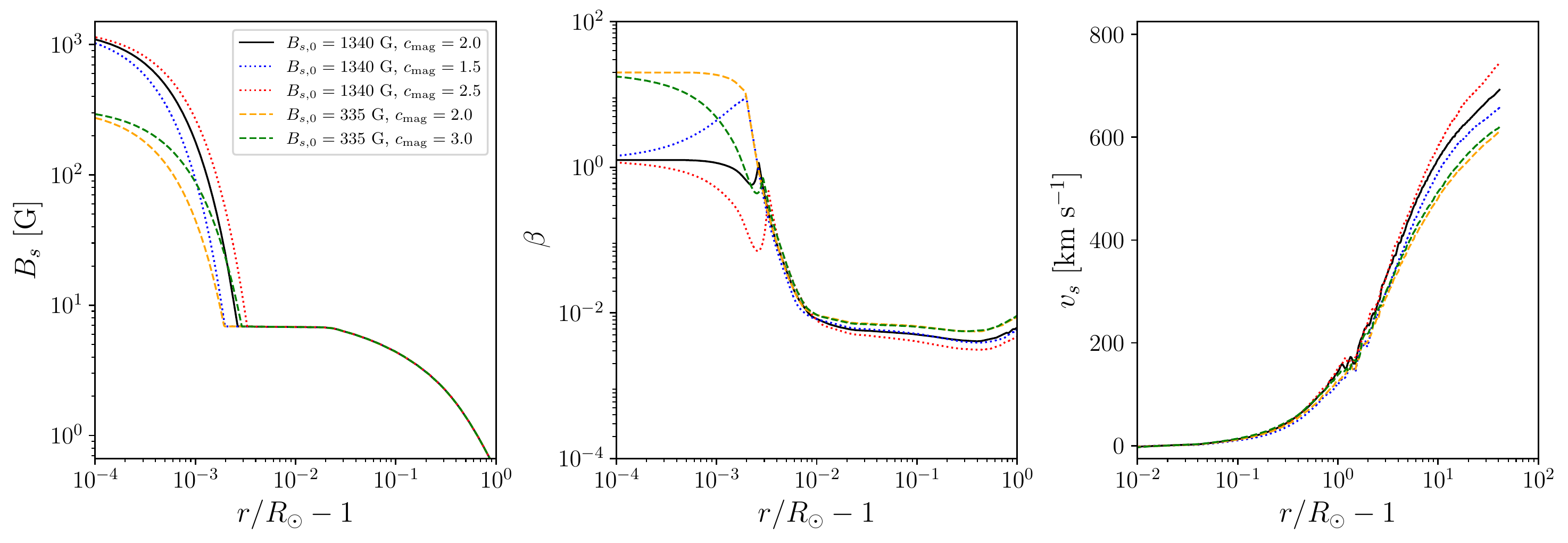}
\vspace{0em}
\caption{
Time-averaged radial profiles of the background magnetic field $B_s$ (left), plasma beta $\beta=8 \pi p/B_s^2$ (middle), and tube-aligned velocity (right).
The black solid line corresponds to the fiducial case ($B_{s,0} = 1340 {\rm \ G}$ and $c_{\rm mag}=2.0$).
The dotted lines show the cases with the same photospheric field strength ($B_{s,0} = 1340 {\rm \ G}$) and the different values of $c_{\rm mag}$ 
(blue: $c_{\rm mag}=1.5$, red: $c_{\rm mag}=2.5$).
The dashed lines show the cases with the smaller photospheric field strength 
(orange: $B_{s,0} = 1340 {\rm \ G}$ and $c_{\rm mag}=2.0$, green: $B_{s,0} = 1340 {\rm \ G}$ and $c_{\rm mag}=3.0$).
}
\label{fig:chrom_mag_survey}
\vspace{2em}
\end{figure*}

Figure ~\ref{fig:chrom_mag_survey} shows the overview of this survey.
The three panels display the time-averaged radial profiles of $B_s$ (left), $\beta=8 \pi p/B_s^2$ (middle), and  $v_s$ (right), respectively.
The black solid line corresponds to the fiducial case ($B_{s,0} = 1340 {\rm \ G}$ and $c_{\rm mag}=2.0$), in which $\beta \approx 1$ in the lower atmosphere ($r/R_\odot-1<10^{-3}$), as seen in the middle panel.
The two dotted lines (red and blue) are the cases with $B_{s,0} = 1340 {\rm \ G}$ but different $c_{\rm mag}$ values.
We note that the blue and red lines correspond to the high- and low-beta chromospheres, respectively.
In spite of substantial differences in the chromospheric magnetic field, the solar wind velocity remains similar.
The two dashed lines (orange and green) are the cases with smaller $B_{s,0}$ ($B_{s,0} = 335 {\rm \ G}$).
Again, although the profiles in the plasma beta is significantly different from the fiducial case, 
the resultant solar wind velocity is similar.
These results indicate that the chromospheric magnetic field has a weak influence on the solar wind velocity.

To see more quantitatively the uncertainty that comes from the chromospheric magnetic field, Table ~\ref{table:appendix_list} shows the values of the solar wind velocity $v_r$ and the mass-loss rate $\dot{M}_w = 4 \pi r^2 \rho v_r$, both measured at $r/R_\odot=30$.
According to the table, the wind velocity typically has an uncertainty of $\pm 80 {\rm \ km \ s^{-1}}$ that comes from the choice of the chromospheric magnetic field.
In terms of mass-loss rate, the uncertainty is as large as $\sim 50 \%$, much larger than that of the wind velocity.
To conclude, the solar wind velocity is weakly affected by the chromospheric magnetic field, while it has a larger impact on the mass flux.

\end{appendix}

\bibliographystyle{aasjournal}

\end{document}